\documentclass[12pt]{iopart}
\usepackage{iopams}
\usepackage{graphicx}
\begin{document}

\title[Traces of Thermalization at RHIC]{Traces of Thermalization at RHIC}

\author{Sean Gavin}

\address{Physics Department, Wayne State
University, Detroit, MI 48201}

\begin{abstract}
I argue that measurements of Au+Au collisions at 20, 130 and 200
GeV of the centrality dependence of the mean $p_t$ together with
$p_t$ and net-charge fluctuations reflect the approach to local
thermal equilibrium.
\end{abstract}




\section{Introduction}


Fluctuations of the net transverse momentum reported by the CERES,
NA49, PHENIX and STAR experiments exhibit substantial dynamic
contributions \cite{Mitchell}. In particular, preliminary STAR
data for Au+Au collisions at 20, 130 and 200~GeV energies show
that $p_t$ fluctuations increase as centrality increases
\cite{Pruneau}. However, data from PHENIX and STAR exhibit a
similar increase in the mean transverse momentum $\langle
p_t\rangle$, a quantity unaffected by fluctuations
\cite{StarMean,Adler:2003cb}.
In \cite{Gavin04} I ask whether the approach to local thermal
equilibrium can explain the similar centrality dependence of
$\langle p_t\rangle$ and $p_t$ fluctuations. Here I address new
data from STAR and PHENIX, including a new and, perhaps, related
effect in the centrality dependence of net-charge fluctuations
\cite{Pruneau}.

Dynamic fluctuations are generally determined from the measured
fluctuations by subtracting the statistical value expected, e.g.,
in equilibrium \cite{PruneauGavinVoloshin}. For particles of
momenta $\mathbf{p}_1$ and $\mathbf{p}_2$, dynamic multiplicity
fluctuations are characterized by
\begin{equation}\label{eq:DynamicMult}
    R_{AA}={{\langle N^2\rangle -\langle N\rangle^2 -\langle N\rangle}\over{\langle
    N\rangle^2}}={{1}\over{\langle N\rangle^{2}}}\int\! d\mathbf{p}_{1}d\mathbf{p}_{2}\,
    r(\mathbf{p}_{1},\mathbf{p}_{2}),
\end{equation}
where $\langle \cdots\rangle$ is the event average. This quantity
depends only on the two-body correlation function
$r(\mathbf{p}_{1},\mathbf{p}_{2}) =
N(\mathbf{p}_{1},\mathbf{p}_{2}) -
N(\mathbf{p}_1)N(\mathbf{p}_2)$. It is obtained from the
multiplicity variance by subtracting its Poisson value $\langle
N\rangle$, and divided by $\langle N\rangle^2$ to minimized the
effect of experimental efficiency \cite{PruneauGavinVoloshin}. For
dynamic $p_t$ fluctuations one similarly finds
\begin{equation}\label{eq:Dynamic}
    \langle \delta p_{t1}\delta p_{t2}\rangle =
    \int\! d\mathbf{p}_{1}d\mathbf{p}_{2}\,
    {{r(\mathbf{p}_{1},\mathbf{p}_{2})}\over{\langle
    N(N-1)\rangle}}
    \delta p_{t1} \delta p_{t2},
\end{equation}
where $\delta p_{ti} = p_{ti}-\langle p_t\rangle$; STAR measures
this observable. The observable measured by PHENIX satisfies
$F_{p_t} \approx N \langle\delta p_{t1}\delta
p_{t2}\rangle/2\sigma^2$ when dynamic fluctuations are small
compared to statistical fluctuations $\sigma^2 = \langle
p_t^2\rangle - \langle p_t\rangle^2$.

\section{Thermalization}

Thermalization occurs as scattering drives the phase space
distribution within a small fluid cell toward a Boltzmann
distribution that varies in space through the temperature
$T(\mathbf{x}, t)$. The time scale for this process is the
relaxation time $\nu^{-1}$. In contrast, density differences
\emph{between} fluid cells must be dispersed by transport from
cell to cell. The time needed for diffusion to disperse a dense
fluid ``clump'' of size $L \sim (|\nabla n|/n)^{-1}$ is $t_{\rm
d}\sim \nu L^2/v_{\rm th}^2$, where $v_{\rm th}\sim 1$ is the
thermal speed of particles. This time can be much larger than
$\nu^{-1}$ for a sufficiently large clumps. Global equilibrium, in
which the system is uniform, can be only obtained for $t \gg
t_{\rm d}$. However, the rapid expansion of the collision system
prevents inhomogeneity from being dispersed prior to freeze out.

Dynamic fluctuations depend on the number of independent particle
``sources.'' This number changes as the system evolves. Initially,
these sources are the independent strings formed as the nuclei
collide. The system is highly correlated along the string, but is
initially uncorrelated in the transverse plane. As local
equilibration proceeds, the clumps become the sources. The
fluctuations at this stage depend on number of clumps as
determined by the clump size, i.e., the correlation length in the
fluid. Dynamic fluctuations would eventually vanish if the system
reaches global equilibrium, where statistical fluctuations are
determined by the total number of particles.

\section{Mean $p_t$ and its Fluctuations}

In \cite{Gavin04} I use the Boltzmann transport equation to show
that thermalization alters the average transverse momentum
following
\begin{equation}\label{eq:meanPt}
    \langle p_t\rangle = \langle p_t\rangle_o S + \langle
    p_t\rangle_e (1-S),
\end{equation}
where $S\equiv e^{-{\mathcal N}}$ is the probability that a
particle escapes the collision volume without scattering. The
initial value $\langle p_t\rangle_{o}$ is determined by the
particle production mechanism. If the number of collisions
$\mathcal{N}$ is small, $S\approx 1 -\mathcal{N}$ implies the
random-walk-like increase of $\langle p_t\rangle$ relative to
$\langle p_t\rangle_{o}$. For a longer-lived system, energy
conservation limits $\langle p_t\rangle$ to a local equilibrium
value $\langle p_t\rangle_e$ fixed by the temperature.

As centrality is increased, the system lifetime increases,
eventually to a point where local equilibrium is reached.
Correspondingly, the survival probability $S$ in (\ref{eq:meanPt})
decreases with increasing centrality. The average $p_t$ peaks for
impact parameters near the point where equilibrium is established.
The behavior in events at centralities beyond that point depends
on how the subsequent hydrodynamic evolution changes $\langle
p_t\rangle_e$ as the lifetime increases. Systems formed in the
most central collisions can experience a cooling that reduces
(\ref{eq:meanPt}) with proper time $\tau$ as $\langle
p_t\rangle_e\propto \tau^{-\gamma}$ \cite{Gavin04}.

\begin{figure}
\begin{center}
\includegraphics[width=2.95in]{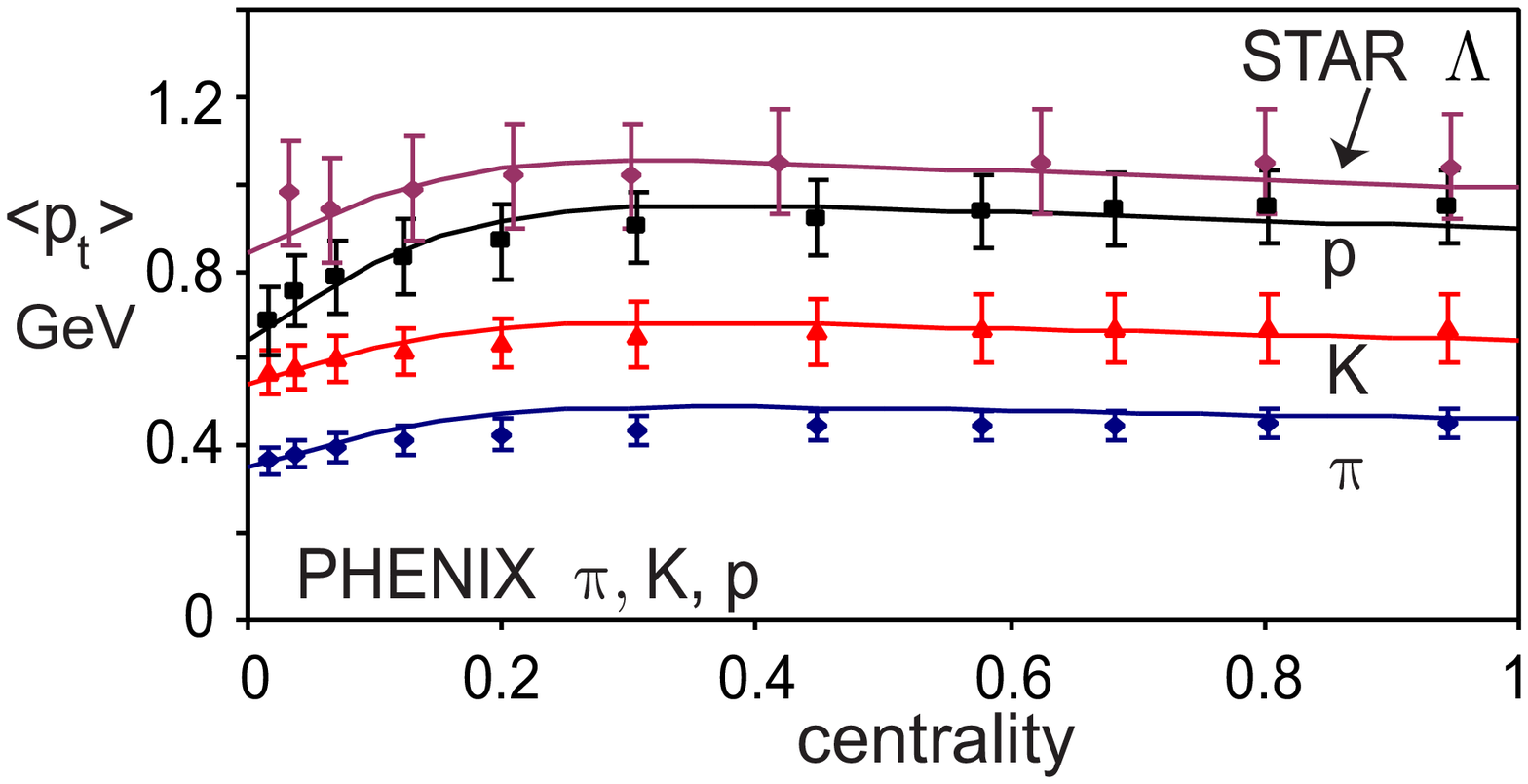}
\includegraphics[width=3.1in]{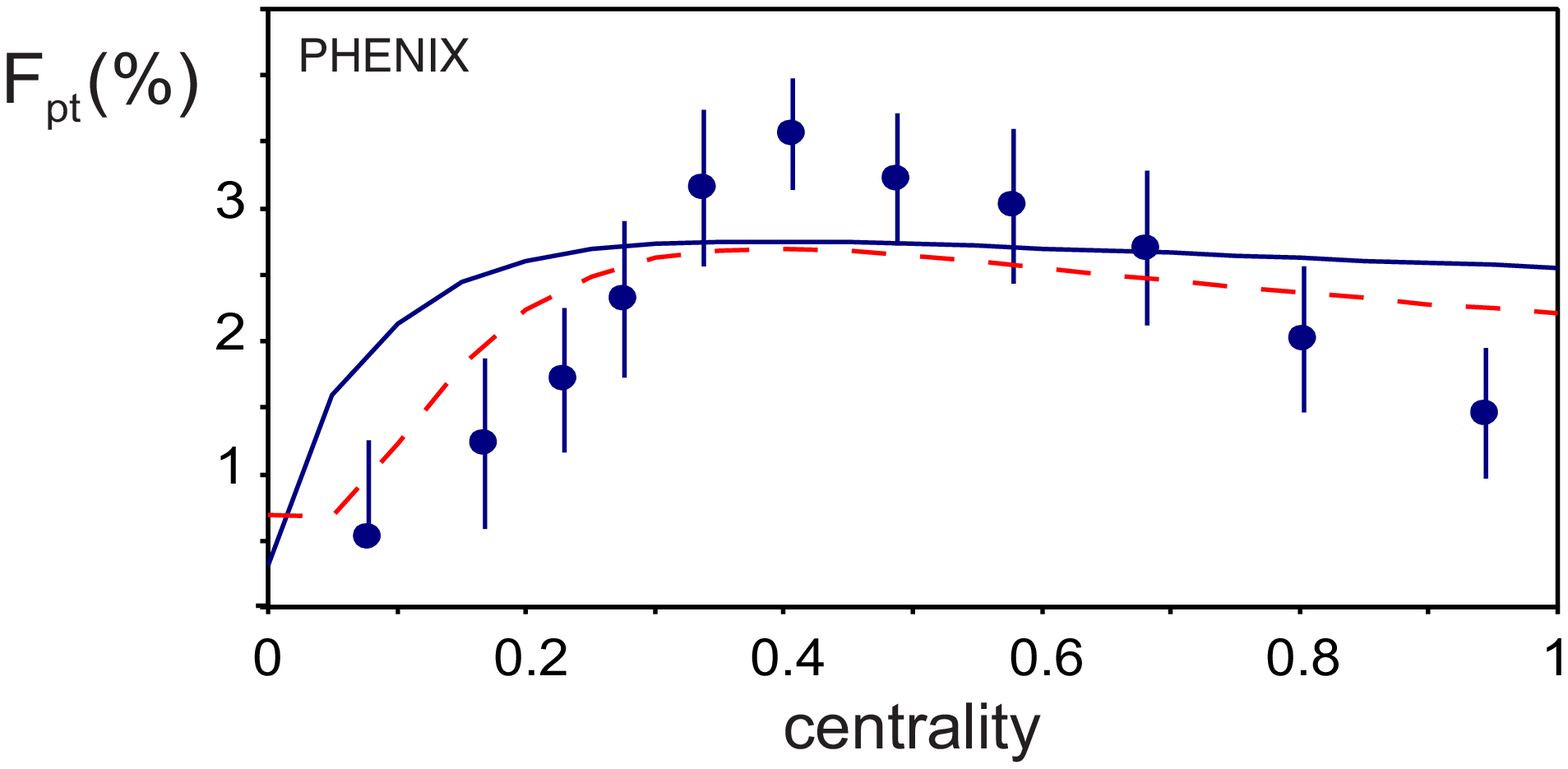}
\caption{Mean $p_t$ (left) and dynamic $p_t$ fluctuations (right)
with data \cite{Mitchell,StarMean,Adler:2003cb}. The dashed curve
in the right figure is a fit to the PHENIX data alone, while the
solid curve is a fit to the STAR data. Centrality is determined by
the number of participants relative to the b=0 value.}
\label{fig:PHENIX}
\end{center}
\end{figure}
Thermalization can explain the behavior in fig.~\ref{fig:PHENIX}.
The survival probability is $S = \exp\{-\int_{\tau_0}^{\tau_F}
\nu(\tau) d\tau\} \approx (\tau_0/\tau_F)^\alpha$, where $\nu =
\langle \sigma v_{\rm rel}\rangle n(\tau)$ for $\sigma$ the
scattering cross section, $v_{\rm rel}$ the relative velocity, and
$\tau_{0,F}$ the formation and freeze out times. Longitudinal
expansion implies $n(\tau)\propto \tau^{-1}$, yielding the power
law with $\alpha = \nu_0\tau_0$. To fit the measured centrality
dependence, I assume $\alpha = 4$ and $\gamma = 0.15$ in central
collisions, and parameterize $S(N_{\rm part})$ by taking $\alpha
\propto N_{\rm part}^{1/2}$ and $\tau_F -\tau_0 \propto N_{\rm
part}$, where $N_{\rm part}$ is the number of participants. I take
the same $\alpha$ for all species, as appropriate for parton
scattering.

Dynamic fluctuations depend on two-body correlations and,
correspondingly, are quadratic in the survival probability. In
\cite{Gavin04} I add Langevin noise terms to the Boltzmann
equation to describe the fluctuations of the phase space
distribution. A simple limit is obtained when the initial
correlations are independent of those near local equilibrium,
    $\langle \delta p_{t1}\delta p_{t2}\rangle
    =\langle \delta p_{t1}\delta p_{t2}\rangle_o S^2
    + \langle \delta p_{t1}\delta p_{t2}\rangle_e (1-S)^2$;
this form was used in \cite{Gavin04}. Alternatively, if the
initial correlations are not far from the local equilibrium value,
I find
\begin{equation}\label{eq:ptFluct}
    \langle \delta p_{t1}\delta p_{t2}\rangle
    =\langle \delta p_{t1}\delta p_{t2}\rangle_o S^2
    + \langle \delta p_{t1}\delta p_{t2}\rangle_e (1-S^2).
\end{equation}
Here I use (\ref{eq:ptFluct}), which provides somewhat better
agreement with the latest STAR data in the peripheral region where
thermalization is incomplete and my model assumptions most
applicable. As before, the initial quantity $\langle \delta
p_{t1}\delta p_{t2}\rangle_o$ is determined by the particle
production mechanism, while $\langle \delta p_{t1}\delta
p_{t2}\rangle_e$ describes the system near local equilibrium.

To estimate $\langle \delta p_{t1}\delta p_{t2}\rangle_o$ for
nuclear collisions, I apply the wounded nucleon model to describe
the soft production that dominates $\langle p_t\rangle$ and
$\langle \delta p_{t1}\delta p_{t2}\rangle$, to find
\begin{equation}\label{eq:wnm}
    \langle \delta p_{t1}\delta p_{t2}\rangle_o =
    {{2\langle \delta p_{t1}\delta p_{t2}\rangle_{pp}}\over{N_{\rm part}}}
    \left({{1+R_{pp}}\over{1+R_{AA}}}\right)
\end{equation}
\cite{Gavin04}. The pre-factor is expected because
(\ref{eq:Dynamic}) measures relative fluctuations and, therefore,
should scale as $N_{\rm part}^{-1}$. The term in parentheses
accounts for the normalization of (\ref{eq:Dynamic}) to $\langle
N(N-1)\rangle$; $R_{AA}$ scales as $N_{\rm part}^{-1}$
\cite{PruneauGavinVoloshin}. ISR measurements imply $\langle
\delta p_{t1}\delta p_{t2}\rangle_{pp}/\langle p_t\rangle_{pp}^2
\approx 0.015$. I use {\textsc{HIJING}} to estimate $R_{pp}$ and
$R_{AA}$, thus building in resonance and jet fluctuations.

Near local equilibrium, spatial correlations occur because the
fluid is inhomogeneous -- it is more likely to find particles near
a dense clump. These spatial correlations fully determine the
momentum correlations since the distribution at each point is
thermal. The mean $p_t$ at each point is proportional to the
temperature $T(\mathbf{x})$, so that
%
    $\langle \delta p_{t1}\delta p_{t2}\rangle_e
    \sim \int 
    r(\mathbf{x}_1,\mathbf{x}_2)
    {\delta T}(\mathbf{x}_1)
    {\delta T}(\mathbf{x}_2)$,
%
where $r(\mathbf{x}_1,\mathbf{x}_2)$ is the spatial correlation
function, $\delta T = T - \overline{T}$, and $\overline{T}$ is a
density-weighted average. I take $n$ ($\propto T^3$) and $r$ to be
Gaussian with the transverse widths $R_t$ and $\xi_t$,
respectively the system radius and correlation length.
In \cite{Gavin04} I obtain
\begin{equation}\label{eq:near}
    \langle \delta p_{t1}\delta p_{t2}\rangle_{e}
    = F{{\langle p_t\rangle^2R_{AA}}\over{ 1+R_{AA}}}
\end{equation}
where $R_{AA}$ is given by (\ref{eq:DynamicMult}) \cite{Gavin04}.
The dimensionless quantity $F$ depends on $\xi_t/R_t$. I compute
$F$ assuming $R_t\propto N_{\rm part}^{1/2}$; $F = 0.046$ for
$\xi_t/R_t = 1/6$. I again use {\textsc{HIJING} to estimate
$R_{AA}$.

\begin{figure}
\begin{center}
\includegraphics[width=2.9in]{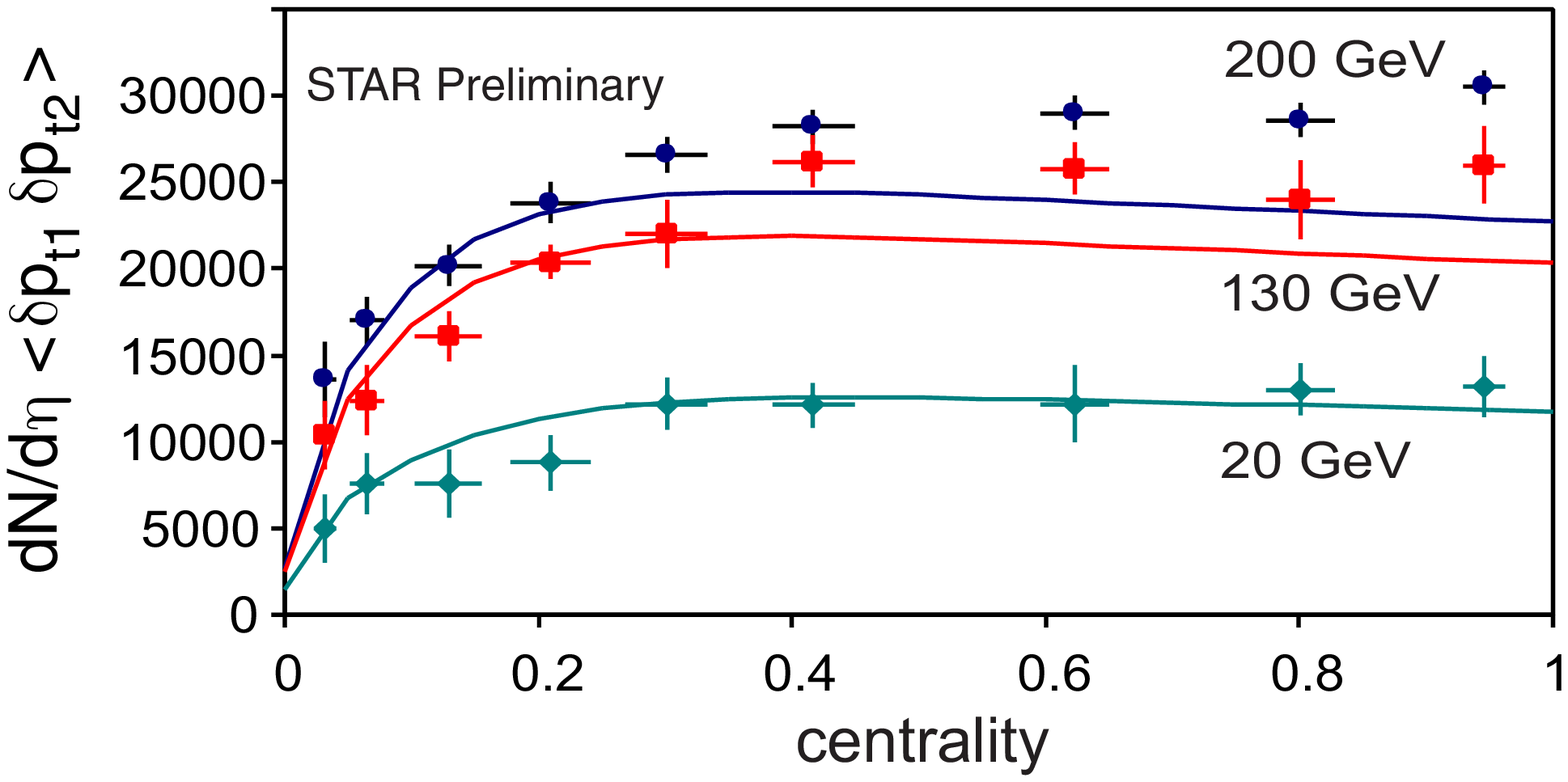}
\includegraphics[width=3.1in]{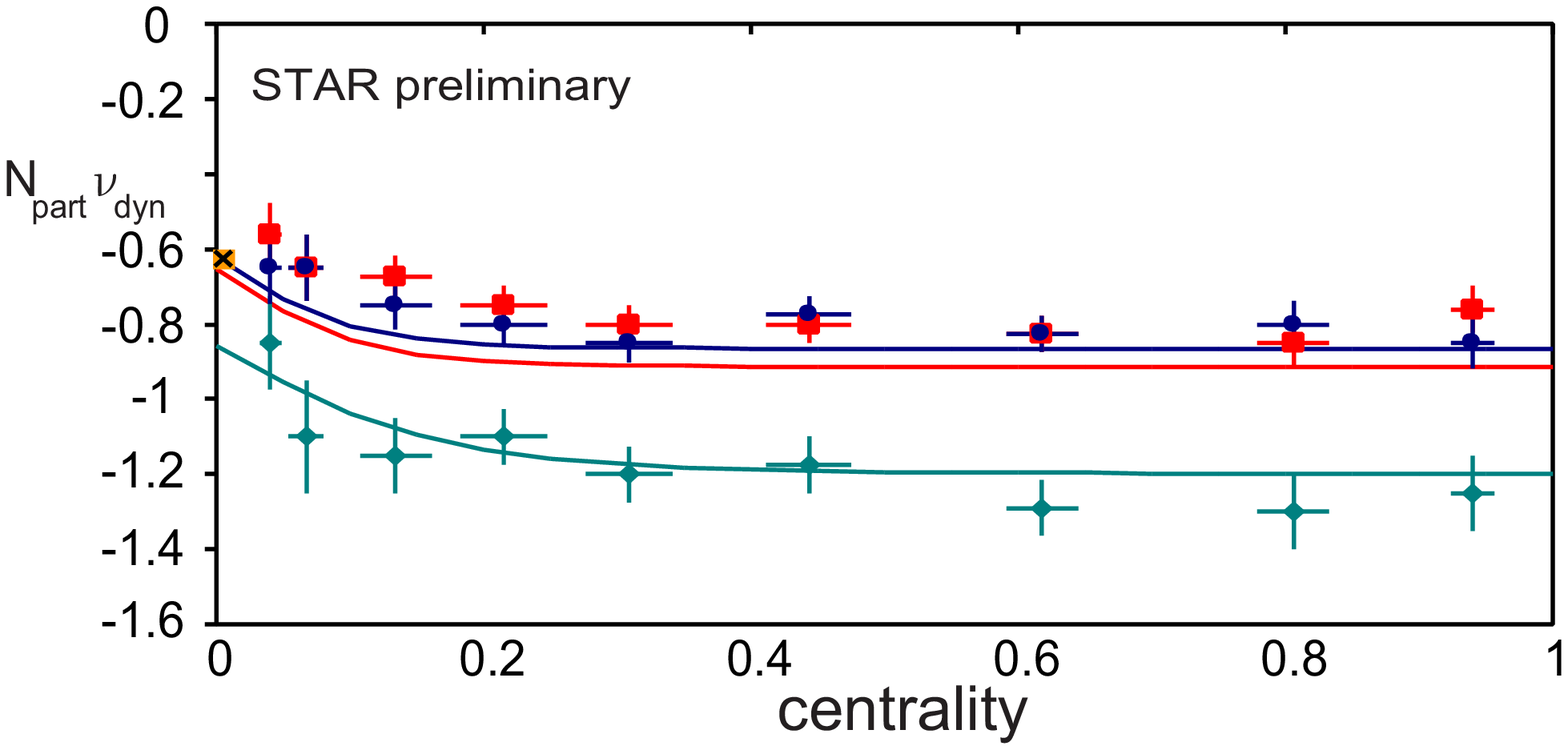}
\caption{Dynamic $p_t$ (left) and net-charge fluctuations (right)
with STAR data \cite{Pruneau}. The curves and data are for three
different beam energies, 20 GeV, 130 GeV and 200 GeV. Centrality
is determined by the number of participants relative to the b=0
value.} \label{fig:ptFluct}
\end{center}
\end{figure}
Calculations in figs.~\ref{fig:PHENIX} and \ref{fig:ptFluct}
illustrate the common effect of thermalization on one-body and
two-body $p_t$ observables. The solid curves in all figures are
fit to STAR fluctuation data and $\langle p_t\rangle$ data, while
the dashed curve shows a fit to the PHENIX data alone. New to this
work are the STAR $p_t$ and net charge fluctuation data in
fig.~\ref{fig:ptFluct}. To compute the energy dependence of
fluctuations, I take $\alpha \propto N$ and $R_{AA}\propto N^{-1}$
for the measured multiplicity $N$. Deviations in the most central
collisions at the highest beam energies may result from radial
flow or, perhaps, jets \cite{Mitchell}. Nevertheless, I stress
that jets -- and quenching effects -- are incorporated in my
calculations via the {\textsc{HIJING}} $R_{AA}$ in
(\ref{eq:near}). {\textsc{HIJING}} without rescattering does not
describe this data.

Net charge fluctuations in fig.~\ref{fig:ptFluct} are
characterized by $\nu_{dyn} = R^{++} + R^{--} -2R^{+-}$ for
$R^{ab}$ given by (\ref{eq:DynamicMult})
\cite{PruneauGavinVoloshin}. I therefore expect $\nu_{dyn}$ to
satisfy $\nu_{dyn} =\nu_o S^2 + \nu_e (1-S^2)$. I use STAR pp data
to fix $\nu_0$ and, following \cite{PruneauGavinVoloshin}, compute
$\nu_e/\nu_0$ assuming a longitudinal narrowing of the correlation
length seen in balance function data \cite{Mitchell}. The
agreement of in-progress calculations with preliminary data
\cite{Pruneau} is encouraging.



\section*{References}
\begin{thebibliography}{99}


\bibitem{Mitchell} J. Mitchell [PHENIX] in these proceedings.
%
\bibitem{Pruneau} G. Westfall [STAR] \emph{ibid.}; C. Pruneau [STAR],
nucl-ex/0401016.
%
\bibitem{StarMean}
J. Adams et al.~[STAR Collab.], nucl-ex/0403020.
%
\bibitem{Adler:2003cb}
S.~S.~Adler  [PHENIX Collab.], nucl-ex/0307022.
%
\bibitem{Gavin04} S. Gavin, Phys.~Rev.~Lett.\, to be
published (2004) [nucl-th/0308067].
%
\bibitem{PruneauGavinVoloshin}
C.~Pruneau, S.~Gavin and S.~Voloshin, Phys.\ Rev.\ C {\bf 66},
044904 (2002) [nucl-ex/0204011].

\end{thebibliography}
\end{document}